\documentclass[%
 reprint,
 amsmath,amssymb,
 aps,
]{revtex4-1}

\usepackage{graphicx}
\usepackage{dcolumn}
\usepackage{bm}
\usepackage[utf8x]{inputenc}
\usepackage{hyperref}

\begin{document}

\title{Magnetic memory effect in ensembles of interacting anisotropic magnetic nanoparticles}
\author{Korobi Konwar$^1$, Som Datta Kaushik$^2$, Debasis Sen$^3$ and Pritam Deb$^1$}
\email{Corresponding Author: pdeb@tezu.ernet.in (Pritam Deb).}
\affiliation{$^1$Department of Physics, Tezpur University (Central University), Tezpur-784028, India.\\
{$^2$} UGC-DAE Consortium for Scientific Research, Mumbai Centre, R-5 Shed, Bhaba Atomic Research Centre, Mumbai 400085, India.\\
{$^3$} Solid State Physics Division, Bhabha Atomic Research Centre, Mumbai 400085, India.}

\date{\today}

\begin{abstract}
We explore the influence of demagnetization interaction on magnetic memory effect by varying organization geometry of anisotropic ZnFe$_2$O$_4$ nanoparticles in an ensemble. The static and dynamic behaviour of two differently organized ensembles, compact ensemble (CE) and hollow core ensemble (HCE), are extensively studied by both dc and ac susceptibility, magnetic memory effect and spin relaxation. The frequency-dependence peak shifting of freezing temperature in both the systems is analyzed properly with the help of two dynamic scaling models: Vogel-Fulcher law and power law. Presence of cluster spin-glass phase is reflected from Vogel-Fulcher temperature $T_0$ $\simeq$ 142.58 K for CE, $\simeq$ 97 K for HCE and characteristic time constant $\tau_0$ $\simeq$ $8.85\times10^{-9}$ s for CE, $\simeq$ $3.8\times10^{-10}$ s for HCE; along with $\delta$T$_{Th}$ $\sim$ 0.1 for CE and 0.2 for HCE. The power law fitting with dynamic exponent $zv'$ = 6.2 $\pm$ 1.1 for CE, 6.3 $\pm$ 0.5 for HCE and single spin flip $\tau^*$ $\simeq$ $7.7\times10^{-11}$ s for CE, $\simeq$ $1.3\times10^{-10}$ s for HCE provide firm confirmation of cluster spin-glass phase. The progressive spin freezing across multiple metastable states with prominent memory effects is reflected in both the systems via nonequilibrium dynamics study. The hollow core geometry with anisotropic nanoparticles on surface with closer proximity leads to complex anisotropy energy landscape with enhanced demagnetizing field resulting highly frustrated surface spins. As a consequence, more prominent magnetic memory effect is observed in HCE with higher activation energy, reduced blocking temperature and enhanced coercivity than that of CE.
\end{abstract}

\maketitle
\section{\textbf{INTRODUCTION}}
Demagnetizing field, originating from the dipolar interaction, plays a crucial role in the ensemble of nanoscaled magnetic nanoparticles due to the dependency of their collective magnetic behaviours on the energy barriers of magnetic anisotropy \cite{1, 2, 3}. When single-domain magnetic nanoparticles congregate with sufficient packing density forming a dense ensemble, dipolar interaction is developed mutually among the particles \cite{4, 5, 6}. Due to the presence of such strong dipolar interaction, their magnetization states segregated by explicit energy barriers, are no more independent resulting modulation in the collective magnetic behaviour \cite{7, 8}. Indeed, in addition to the dipolar interaction, the randomly oriented anisotropy axes results competition among the spins \cite{9, 10, 11}. In this regard, the collective freezing of the moments along with their arbitrary direction leads to a disordered magnetic state below a certain temperature known as super-spin glass (SSG) state \cite{12, 13, 14, 15}. As a result, non-ergodic magnetically frustrated state arises due to the crossover from blocking of individual spin to the collective spin-freezing\cite{16, 17}. Below such freezing temperature, the system starts to maintain an out-of-equilibrium dynamics with cooling due to the continual slowing of internal motion. This resultis an incapability to reach thermodynamic equilibrium because of the macroscopic equilibration period \cite{18, 19}. Such non-equilibrium spin-glass states show typical characteristics known as magnetic memory effect, aging and rejuvenation \cite{20, 21, 22, 23, 24, 25, 26, 27}, which rely on the concentration of nanoparticles along with the interface exchange coupling\cite{28, 29, 30, 31, 32} and their respective dipolar strength.

SSG state is extensively understood in some alloys, systems with geometrically frustrated lattices, systems having competiting spins due to the antiferromagnetic (AFM)-ferromagnetic(FM) interaction, and in some magnetic impurity induced nobel metals \cite{22, 33, 34, 35, 36, 37, 38, 39, 40, 41, 42}. In spite of such in depth study on Spin-Glass (SG) systems, there is still lack of moment relaxation is described traditionally by Dormann-Bessaic-Fiorani (DBF)\cite{43, 44} and Morup-Tronc (MT)\cite{45} models. The first model predicts the slower relaxation of moments with enhanced dipolar-interaction, whereas MT model believes faster relaxation with increasing dipolar strength. These contradictory hypotheses provide ambiguity regarding interparticle interaction in ensemble of nanoparticles. Such obscurity residing among the dipolar interaction, superparamagnetic and spin-glassy states need to be addressed. In contrast to these models, we consider here a new approach which relies on the impact of geometrical arrangement of the spins along with the distribution of easy axes by considering frustrated geomery of spinel Zinc Ferrite systems. The interplay with relative alignment of their domains and respective geometrical organization can tune the anisotropy and competition among the moments. Moreover, the effect of demagnetizing field on the modification of anisotropic energy landscape is understood to correlate its impact on the moment relaxation and corresponding enhancement in their magnetic memory effect.

In the present study, we demonstrate the demagnetizing field dependent magnetic memory effect in three dimensional ensembles of anisotropic ZnFe$_2$O$_4$ nanoparticles with varied geometry organization and optimized spatial arrangement. Initially, a compact ensemble (CE) of anisotropic nanoparticles is considered for the study. Later on, we reasonably manipulate the degree of frustration by introducing a hollow core ensemble (HCE) of dipolar-interacting spins. The non-equilibrium dynamics is extensively evaluated with both the DC and AC magnetization analyses which further confirm the presence of cluster-SG phase in both CE and HCE. By considering the geometry of the ensembles and alignment of easy axes along with strength of demagnetization field, the respective modification in the magnetic memory effect in CE and HCE is studied.

\section{\textbf{EXPERIMENAL DETAILS}}
The core of the synthesis procedure of compact ensemble (CE) and Hollow Core Ensemble (HCE) of $ZnFe_2O_4$ anisotropic nanoparticles is followed by a modified template free solvothermal synthesis protocol \cite{46}. All the precursors required for the synthesis are procured from Zenith, India which are analytical grade and use directly without further sanitisation. The stoichiometric amounts of Zinc acetate dihydrate and Ferric nitrate nonahydrate are keeping under magnetic stirring in a solution of appropriate amount of glycerol and isopropyl alcohol. The reddish homogeneous solution is autoclaved for 12 hours and 21 hours respectively at 180$^\circ$ C. The collected greenish-yellow product followed by centrifugation and drying is calcined at 400$^\circ$ C for 2 hours and dark brown coloured powdered sample is obtained. Finally compact ensemble (CE) and hollow core ensemble (HCE) of anisotropic Zinc Ferrite nanoparticles is obtained. Powder X-ray Diffraction (PXRD) characterization is performed (Rigaku diffractometer with radiation source Cu-$K_\alpha$ at a scan rate of 1$^\circ$/min). High Resolution Transmission Electron Microscopy (HRTEM) is performed on a model JEOL, JEM-2100 PlusElectron Microscope with operating voltage of 200 KV. Small Angle X-ray Scattering (SAXS) technique ie employed in a point collimated SAXS instrument bearing wavelength of X-ray ($\lambda$) $\sim$ 0.154 nm. Double crystal based medium resolution small angle neutron scattering (MSANS) experiment is performed with neutron wavelength $\sim$ 0.312 nm. The magnetic property analyses are carried out in a Vibrating Sample Magnetometer (VSM) of Quantum Design Dynacool Physical Property Measurement System (PPMS) having magnetic field range -9T to +9T.

\begin{figure}[th!]     
\centering           
\includegraphics[width=8.5cm,height=12cm]{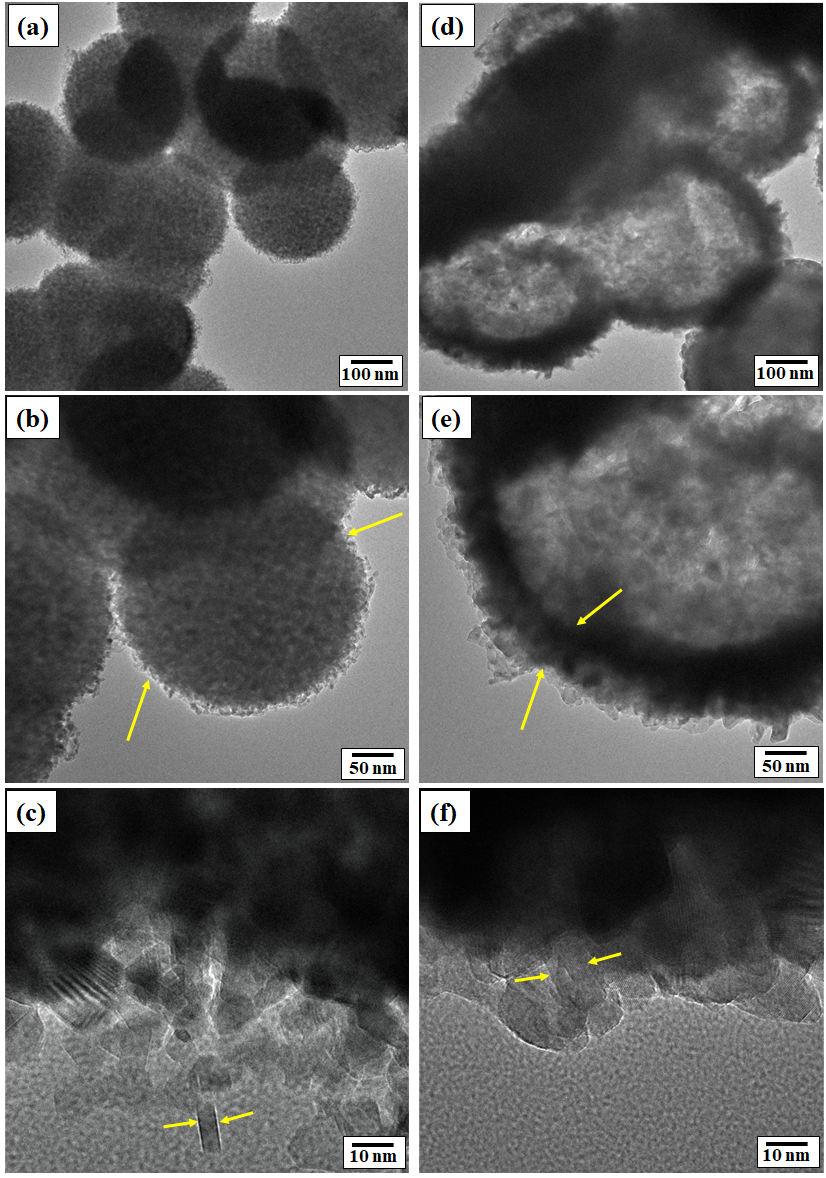}
\caption{\label{Fig:wide} HRTEM micrographs of (a, b, c) CE and (d, e, f) HCE}
\end{figure}

\begin{figure*}[th!]     
\centering           
\includegraphics[width=16cm,height=8cm]{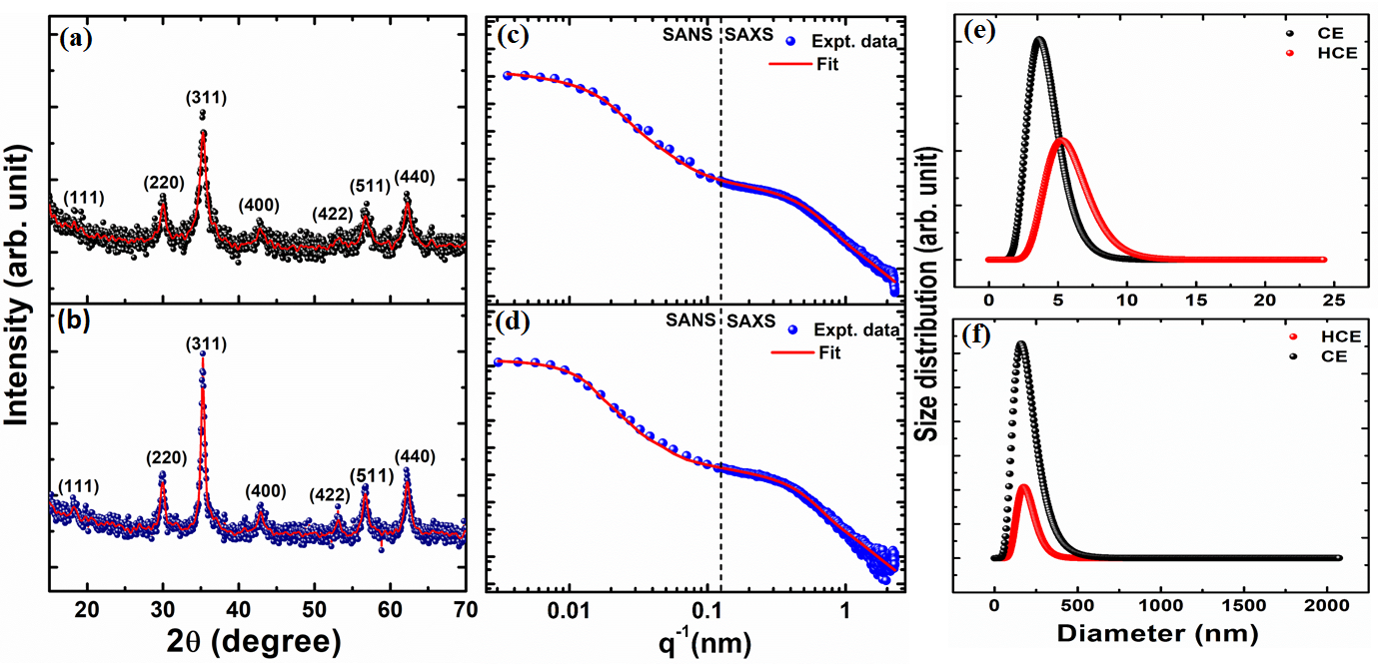}
\caption{\label{Fig:wide} PXRD pattern of (a) CE, (b) HCE; SAXS and MSANS intensity profile with log-Norm fitting by using SASfit software package (c)CE, (d)HCE; particle size distribution of both CE and HCE (e) primary nanoparticles from SAXS, (f) secondary ensembles from SANS}
\end{figure*}

\section{\textbf{RESULS AND DISCUSSIONS}}
\subsection{\textbf{X- ray diffraction and Microstructural study}}

Fig. 1 shows HRTEM micrographs manifesting the formation of isotropic ensembles with anisotropic nanoparticles having varied organization pattern. As shown in figure 1(a, b, c), anisotropic nanoparticles having average size 4$\pm$2 nm are assembled with some interparticle space making a compact ensemble (CE) with an average size of 270 nm. With increase in solvothermal reaction time, the nanoparticles start to accumulate on the surface with lower defects resulting an ensemble of hollow interior (named as HCE) with larger domain size following 'Inside-out Ostwald ripening' mechanism\cite{47}. Figure 1(d, e, f) show the hollow core ensemble (with average size of anisotropic nanoparticles 6$\pm$2 nm and ensemble 270 nm). The accumulation of nanoparticles on the surafce is ensured from the dark contrast marked with the yellow arrows in figure 1 (e) and the light contrast area indicates the hollow interior.
Figure 2 (a), 2(b) show the XRD analysis to confirm the structural phase and crystalline property for both the systems. The observed diffraction peaks at (220), (311), (400), (511) and (440) in the XRD patterns are matching to the cubic spinel phase of ZnFe$_2$O$_4$ (JCPDS card no. 82-1042) ensuring the retaining of Zinc Ferrite phase in both the ensembles CE and HCE \cite{46}. An important point is observed that, with increasing the reaction time to obtain HCE, the crystallinity of the system also get increased. When reaction time is increased, sufficient time is available for nucleation and growth process, which enables packing of particles and the system becomes more ordered.
Figure 2 (c, d) show SAXS and MSANS plots for CE and HCE to obtain overall microstructural information which are supporting the presence of two types of hierarchy. The intensity profile of both the systems obtained from SAXS are fitted properly with scattering model for polydispersive interacting cylindrical system which gives structural information of the primary nanoparticles \cite{48}. The distribution function for particle size is considered in the present scenario as log-normal for both the SAXS and MSANS profile fitting. The SAXS intensity profiles are fitted with form factor of a cylindrical scattered with wave vector q, first order Bessel function $J_1$, radius R and length L respectively and can be represented in terms of independent variable x (limit from 0 to 1) as,
\begin{equation}
\begin{split}
P(q, R, L) = 2\int_{0}^{1} \frac{J_1^2}{\left[qR{(1-x^2)}^\frac{1}{2}\right]}\cdot \frac{\left[1-cos2(\frac{qLx}{2})\right]}{{(\frac{qLx}{2})}^2} dx
\end{split}
\end{equation}

The higher scattering intensity observed in the lower q region for both CE and HCE as shown in figure 2(c, d) resemble the agglomeration behaviour which may be due to annealing induced agglomeration. The higher q region in the SAXS intensity profiles provide information regarding the nature of interaction among the basic particles and can be determined with the help of structure factor. In current scenario, scattering intensity profile of the cylindrical primary particles are modelled by sticky hard sphere type structure factor for both CE and HCE with varied value of the fitting parameters.

Moreover, the SANS intensity profiles (shown in figure (c, d)) are fitted for spherical type form factor,

\begin{equation}
P(q,R) = 4R^3 \pi \eta \left[\frac{sin(qR)-qR\,cos(qR)}{(qR)^3}\right],
\end{equation}

\begin{table*}
\caption{Fitting parameters obtained from SAXS and SANS profile}
\begin{tabular}{ccccccc}
\hline
\multicolumn{1}{c|}{System} & \multicolumn{4}{c|}{Details of Primary nanoparticles from SAXS fitting} & \multicolumn{2}{c|}{Details of Secondary system from MSANS fitting}\\
\hline
{|}Name &{|}Polydispersity Index & {|}Diameter(nm)        & {|}Length(nm) &{|}Interparticle Space(nm)  &{|}Diameter (nm) &{|}Polydispersity Index\\
\hline
CE & 0.22 & 4.2 & 15.98  & 2.5   &190.2 &0.23\\

HCE & 0.24 & 5.6  & 19.5 & 1  &193.5 &0.25\\
\hline

\end{tabular}
\end{table*}
After proper fitting, the morphology obtained for the primary nanoparticles (obtained cylindrical from SAS fitting) and secondary particles are supporting the HRTEM results. The particle size distribution curve obtained from SASfit for both the primary particles and secondary ensembles are depticted in figure 2 (e, f). Nevertheless, the slight difference in the mean particle size and interparticle distance can be ascribed by considering that the small angle scattering technique provides information in overall length scale, whereas TEM provides information of a selected length scale only. The scattering data ensures the presence of sticky hard sphere type of interaction along with packing fraction ($\phi$) $0.20$ and $0.32$ for CE and HCE respectively. The other fitting parameters obtained from the respective fitting is depicted in the table 1. The lower value of stickiness $\tau$ (for CE $\sim$ 0.09 and for HCE $\sim$ 0.07) obtained from the structure factor approximation signifies that the system is very compact.

\begin{figure}[t]    
\centering           
\includegraphics[width=8.6cm,height=15cm]{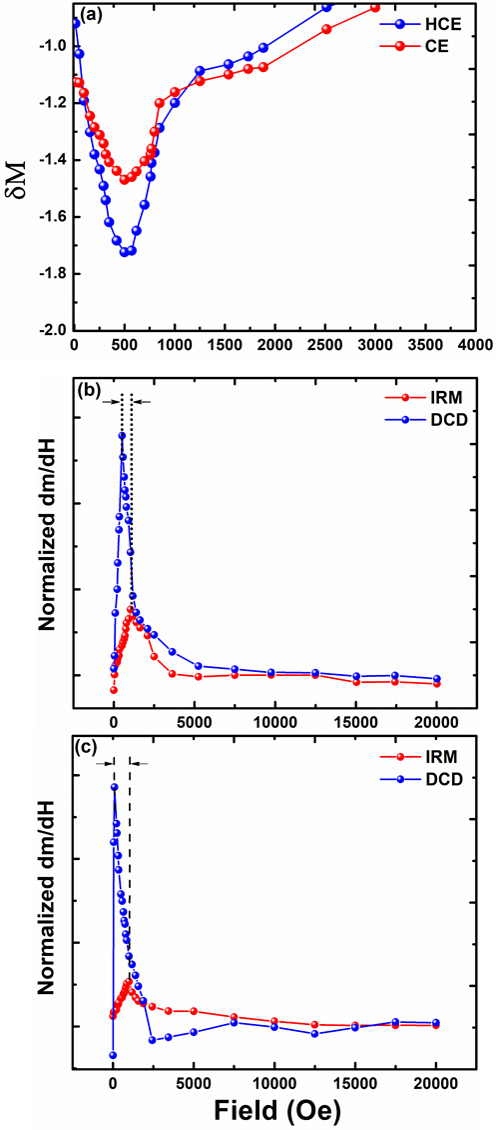}
\caption{\label{Fig:wide} (a)$\delta M$ plots of CE and HCE systems, irreversible susceptibility plot for (b) CE, (c) HCE}
\end{figure}

\subsection{\textbf{Magnetic interaction}}
To reveal the essence of interaction in the ensembles, we analyze the remanence curves which depend on rotation of irreversible magnetization. It provides evidence related to interaction by using $\delta$M plots which is obtained from Direct Current Demagnetization (DCD) as well as Isothermal Remanent Magnetization (IRM) curves. For single domain interacting system with uniaxial anisotropy, IRM and DCD are correlated using Stoner-Wohlfarth expression \cite{49, 1, 2},

\begin{equation}
\delta M = m_{DCD} (H) - [1- 2 m_{IRM} (H)],
\end{equation}

where $m_{DCD}$ and $m_{IRM}$ stand for reduced magnetization of DCD and IRM  respectively. The observed negative deviation of $\delta$M plots as shown in Fig. 3 for both CE and HCE systems reveal the domination of dipolar type of interaction among the primary nanoparticles of the ensembles \cite{1,2}. In addition, differentiation of normalized DCD and IRM curves are plotted to analyze distribution of energy barrier. Considering both the remanence curves, irreversible susceptibility can be compared as, $\frac{\partial m^{DCD}}{\partial H}=2 \frac{\partial m^{IRM}}{\partial H}$. The magnitude of interaction field $H_{int}$ can be expressed as follows, $H_{int}=\frac{1}{2}(H_r - H_r^*)$, where, H$_r$ and H$_r^*$ refer to peak position of field derivative of moments for IRM and DCD respectively. The calculated values of H$_{int}$ are found as -0.25 kOe and -0.45 kOe for CE and HCE respectively. The negative value indicates pre-domination of demagnetizating interaction corroborating negative deviation observed in $\delta$M plots. The higher magnitude of H$_{int}$ for HCE ensures higher demagnetizating interaction than CE. Nevertheless, the probable contribution of exchange interaction in such packed ensembles of nanoparticles cannot be ignored.

\subsection{\textbf{DC magnetization}}
Fig. 4 shows magnetic field dependent magnetization (M-H) trend with variation in temperature (300 K, 150 K, 30 K and 5 K)for both CE and HCE. At room temperature, the magnetic isotherm follows a superparamagnetic pattern with narrow hysteresis nature \cite{6}. With decrease in temperature, the coercivity and remanence show an usual increasing trend and at 5 Oe it shows a coercive field $\simeq$ 690 Oe for the system CE and $\simeq$ 759 Oe for the system HCE respectively. Higher value in coercivity (H$_c$) is observed in HCE than that of CE (at room temperature, the value is  $H_c$ $\sim$ 37.87 Oe for HCE and $H_c$ $\sim$ 10.59 Oe for CE) even after bearing higher dipolar interaction. It is attributed due to the partially alligned easy axes with complex anisotropy landscape. Conventionally, the strong demagnetizing effect helps for making easy the magnetic reversal act \cite{50, 2}. But, here the enhanced dipolar interaction helps to give rise to enhance in coercivity. In essence, the strong demagnetization strength present in the system leads to intense energy barriers in the complex interacting energy landscape. It results the attractive configurations being more attractive in nature. In such case, the required thermal energy to overcome these complex landscape energy barrier needs to be enhanced. If the thermal energy is not sufficient enough, it would be less probable to overcome the energy barriers resulting enhancement of coercivity in the system.

\begin{figure}[th!]     
\centering           
\includegraphics[width=8.6 cm,height=14 cm]{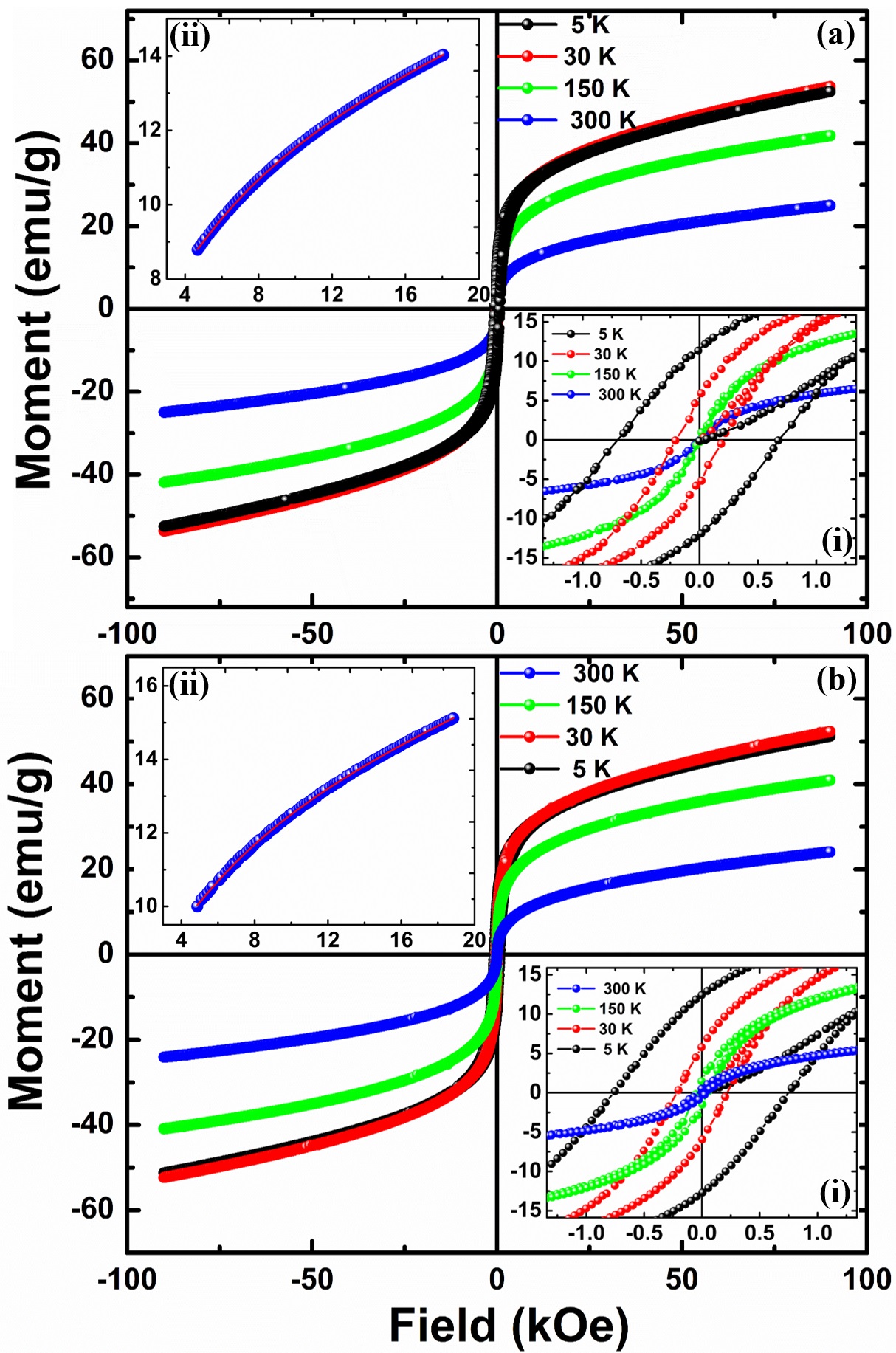}
\caption{\label{Fig:wide} Probe field relying magnetization (a)CE and (b)HCE, inset shows zoomed field relying magnetization (a)(i)CE, b (ii)HCE, LAS fitting of (a)(ii)HC, (b) (ii)HCE}
\end{figure}

\begin{figure}[t]     
\centering           
\includegraphics[width=8 cm,height=12 cm]{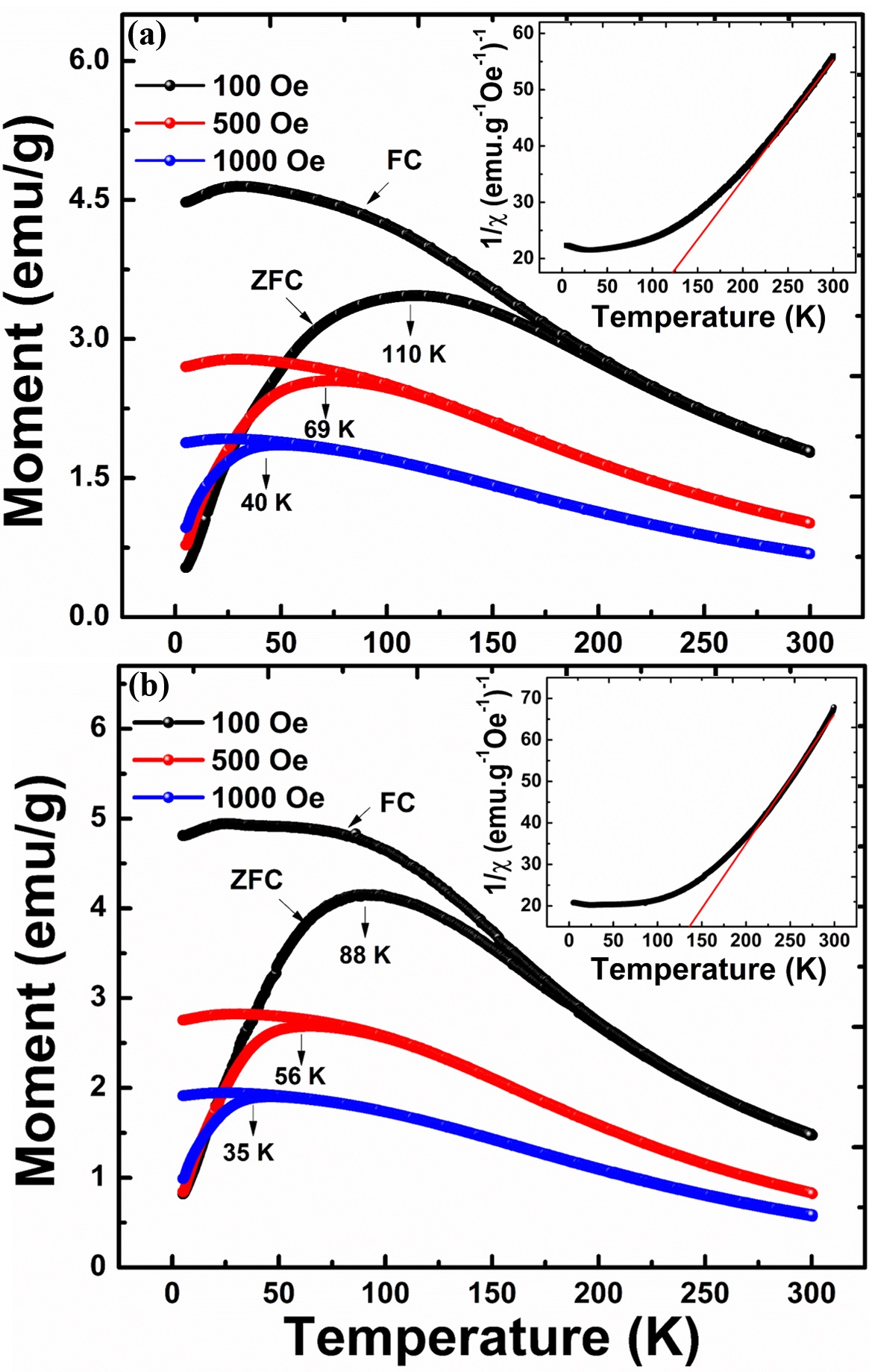}
\caption{\label Temperature relying magnetization plot (a)CE and (b) HCE; Curie-Weiss (WS) law fitting in inset}
\end{figure}

The saturation magnetization and anisotropy constant are measured by fitting experimental data of magnetization for field range 4kOe$<$H$<$18kOe shown in inset of Fig. 4(a(ii), b(ii)) using the law of approach to saturation (LAS) \cite{51} where spins rotate against anisotropy by considering internal demagnetizing field and all decoupled oriented grains,

\begin{equation}
H = H_s \left[ 1- \frac{A}{H} - \frac {B}{H^2}\right] + \kappa H,
\end{equation}
where H is the applied field, $\frac{A}{H}$ can be defined as magnetic hardness which  ascribes the structural defect. The term $\frac{B}{H^2}$ attributes the magnetocrystalline anisotropy which is related to cubic anisotropy constant K and permeability of free space $\mu_0$ as, B = $\frac{8}{105}$$\left[\frac{K^2}{\mu_0^2 M_s^2}\right]$. $\kappa$ represents the forced magnetization which is resulting from increased spontaneous magnetization with high applied magnetic field. The calculated value of anisotropy constant of HCE and CE from the fitted parameters values are found as $83.13\times10^3$ and $87.76\times10^3$ respectively. Moreover, the calculated reduced remanence value for CE and HCE are 0.0008 and 0.04 emu/g, which are significantly less than the theoretical value 0.5 ensuring the presence of single domain nanoparticles with uniaxial anisotropy \cite{1, 2}. The value of reduced remanence can be analysed from the consequence of competition arises between intraparticle anisotropy and dominant demagnetizing interaction on the process of spin relaxation which results frustration.

\begin{figure}[th!]     
\centering           
\includegraphics[width=9cm,height=15cm]{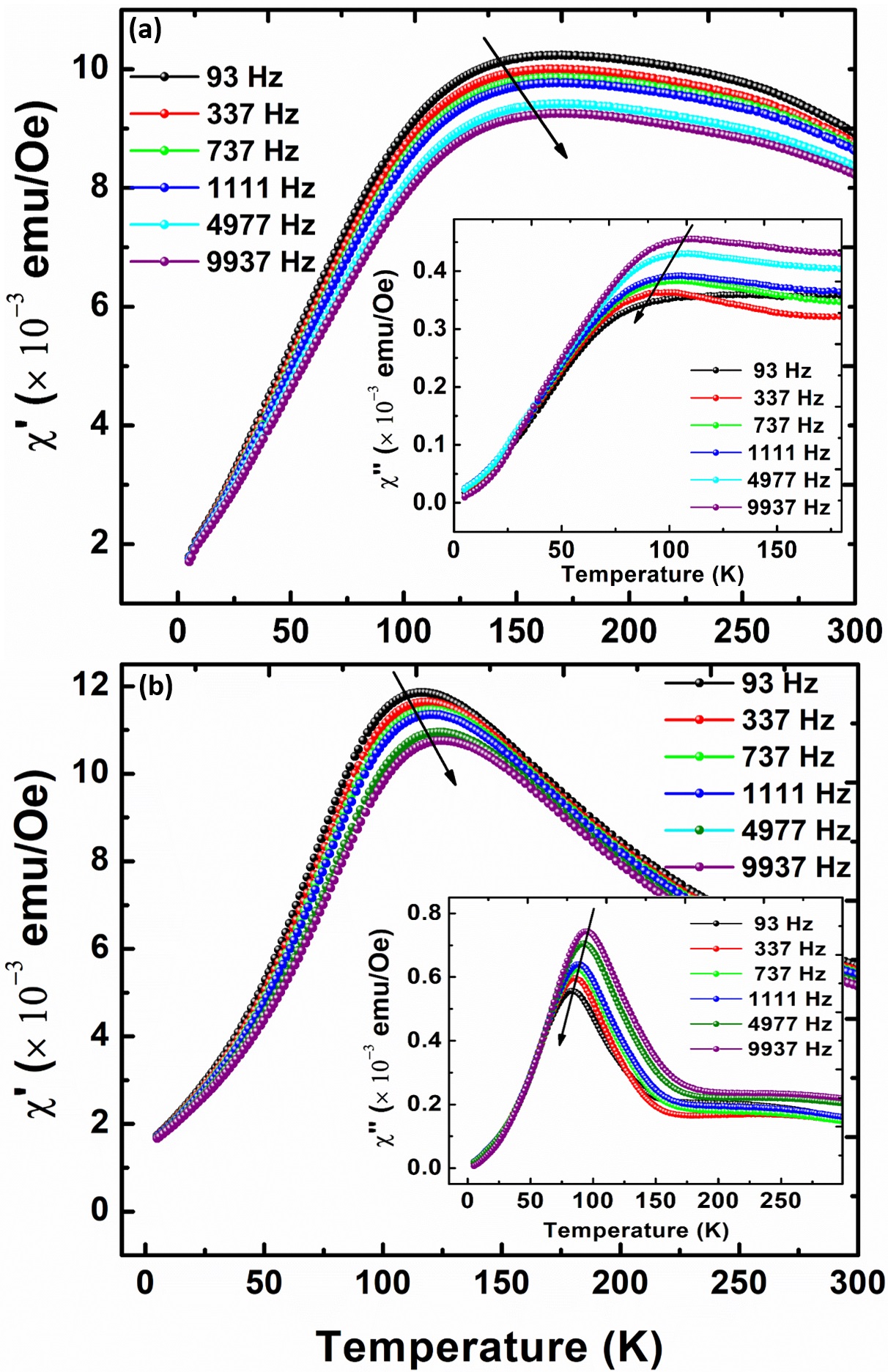}
\caption{\label{Fig:wide} In-phase ac susceptibility at applied field $H_DC$ = 0 kOe with $H_AC$ = 10 Oe, out-of-phase ac susceptibility in inset (a)CE and (b)HCE}
\end{figure}

Fig. 5 presents temperature dependent dc magnetization plots with different magnetic field (1000 Oe, 500 Oe and 100 Oe) using Zero-Field Cooling (ZFC) and Field Cooling (FC) conditions. The observed bifurcation below room temperature implies the existence of superparamagnetic nature for both the systems CE and HCE \cite{2}. The observed broad peak (blocking temperature, $T_B$) in ZFC at around 110 K for CE and 88 K for HCE at 100 Oe are due to the transition of superparamagnetic state to blocked state. To understand the nature of transition, magnetization is measured with different magnetic field. With increase in magnetic field, moments start to decraese leading to lowering in susceptibility along with a plateau in the ZFC data. Subsequently, $T_B$ shifts toward lower value. With enhanced magnetic field, the crystal-field anisotropy decreases and very less amount of thermal energy is required to cross the height of the energy barrier between the two easy axes orientation. When magnetic field is low, the Zeeman energy is smaller than that of the thermal energy and hence the thermal energy plays the dominant role resulting higher value of blocking temperature. Moreover, the dependency of blocking temperature on the dipolar strength is showing an unusual behavior, as the system HCE (having higher dipolar strength than the system CE) is resulting lower blocking temperature violating the DBF model \cite{43, 44}. This may be due to the rapid spin relaxation compelled by the higher demagnetization effect. Nevertheless, in FC curves saturation in magnetization is observed below $T_B$ indicating the presence of strong interaction among the grains in the ensembles \cite{1}. Addionally, with increase in applied field, the difference between the moments consequently the susceptibility ($\Delta$ $\chi$) gets decreased at low temperature in both CE and HCE systems \cite{52}. The shifting in $T_B$, lowering in $\Delta$ $\chi$ and trend of FC curve below $T_B$ at low temperature indicate the possibility of presence of spin-glassy transition at low temperature.

The degree of alignment of easy axes at low temperature region can also be evaluated by low temperature FC magnetization value \cite{2} as,
\begin{equation}
\frac{M_{FC} ^ {aligh}}{M_{FC}} =1 + \alpha (3 \cos ^ 2 \beta -1),
\end{equation}

Here, M$_{FC}^{align}$ defines FC magnetization at a very low temperature (here, we consider 10 K)of a system having aligned easy axes and M$_{FC}$ represents FC magnetization of a system having randomly oriented easy axes. $\alpha$ gives the measure of fraction value of nanoparticles bearing orientated magnetic easy axis. $\beta$ gives the measure of average angle between the easy axes and the magnetic field. If the system contains highly random easy axes, then $\alpha$ yields 0 resulting $\frac{M_{FC} ^ {aligh}}{M_{FC}}$ = 1. Whereas, presence of alignment in the easy axes leads to non-unity value of $\frac{M_{FC} ^ {aligh}}{M_{FC}}$. Initially, we calculate $\frac{M_{FC} ^ {aligh}}{M_{FC}}$ by considering HCE as a system having aligned easy axes and CE as a system of random easy axes. The moment at 10 K for the system HCE and CE are considered as M$_{FC}^{align}$ and M$_{FC}$. The obtained value 1.05 which is nearly tending to unity signifying presence of alignment of easy axes in the system HCE \cite{2}.

As shown in inset of Fig. 5, the inverse susceptibility $\frac{1}{\chi}$ is fitted with the help of Curie-Weiss (CW) law \cite{51, 2} in the high temperature region at applied field 100 Oe,
\begin{equation}
\chi =\frac{C}{(T-\theta_{CW})},
\end{equation}

where C stands for Curie constant and $\theta_{CW}$ represnts the CW temperature respcetively. The best fitting of CW law gives C $\approx$ 4.7 g.cm$^{-3}$K and $\theta_{CW}$ $\approx$ 40 K for the system CE; for HCE C $\approx$ 3.2 g.cm$^{-3}$K and $\theta_{CW}$ $\approx$ 88 K. The presence of positive $\theta_{CW}$ manifests the domination of ferromagnetic ordering in both the systems. The effective magnetic moment is calculated from the obtained C as, $\mu_{eff}$ = $\sqrt{\frac{3k_BC}{N_A}}$ ($N_A$ stands for Avogadro's number). For system CE, $\mu_{eff}$ $\sim$ 6.13 $\mu_B$ and for HCE $\mu_{eff}$ $\sim$ 5.06 $\mu_B$. 

\begin{figure}[t]     
\centering           
\includegraphics[width= 8.6 cm,height= 13 cm]{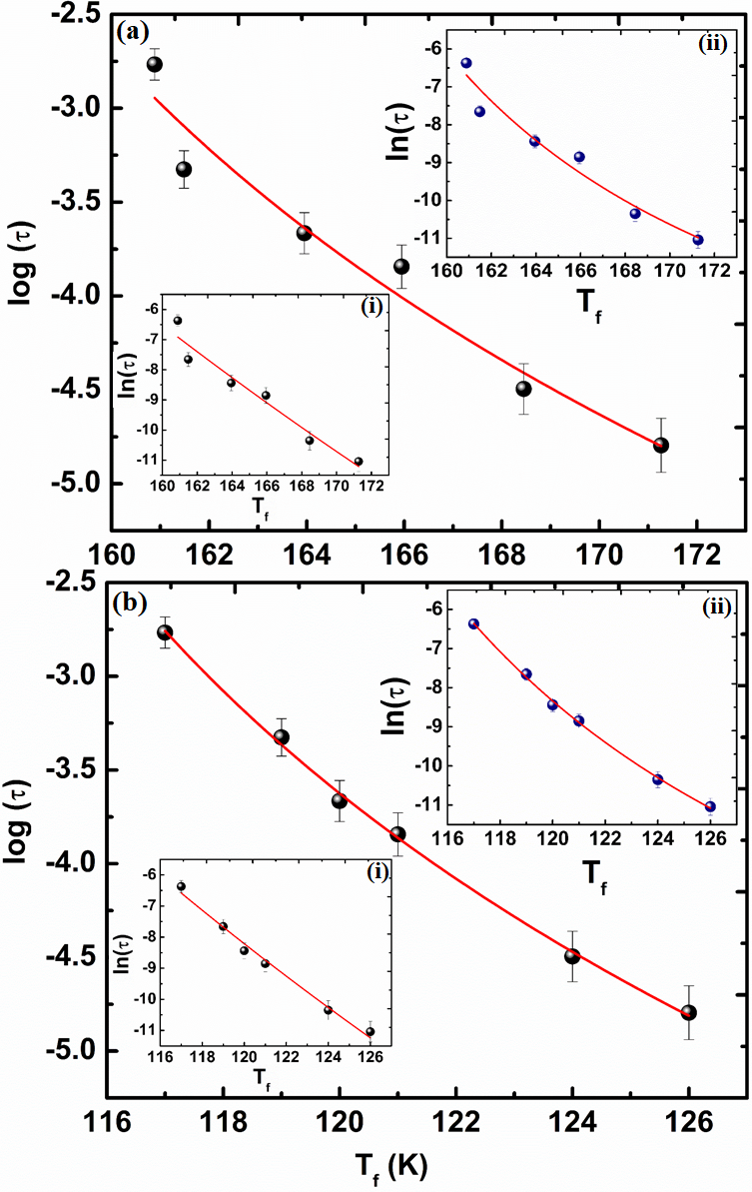}
\caption{\label{Fig:wide} Fitting of critical slowing down model, (i)Arrhenius law fitting in inset (ii)VF law fitting in inset (a)CE and (b)HCE}
\end{figure}

\subsection{\textbf{AC susceptibility}}
In order to verify the insight understanding of the transition nature and detail spin dynamics, ac susceptibility study is systematically executed for both the systems within the frequency range of 93 Hz to 9937 Hz with a constant excitation ac applied field of 10 Oe. Fig. 6 shows both the in-phase and out-of-phase ac susceptibility as a function of temperature for both CE and HCE systems. A frequency dependent pronounced anomaly is observed at around 160.9 K and 117 K (at frequency 93Hz, in-phase ac susceptibility) for the system CE and HCE respectively. The observed peak slightly changes its position towards higher temperature with increase in frequency along with the decrease in peak height which signifies the glassy transition of the systems at T$_f$ $\approx$ 160.9 K for CE and T$_f$ $\approx$ 117 K for HCE \cite{39, 38, 52, 1}. The study is also performed to the system HCE by adding DC field along with the fixed AC field.

The observed peak shift in freezing temperature with varied frequency is evaluated by a phenomenological parameter, known as Mydosh parameter\cite{53} to differentiate among the SG systems and it can be calculated using the relation \cite{1, 52, 39},
\begin{equation}
k = \Delta T_{f}/T_{f}(\Delta\log_{10} f),
\end{equation}

The value is calculated using the outermost frequencies, $\nu_{1}$ = 93 Hz and $\nu_{2}$ = 9937 Hz. $k$ shows a closer value to 0.1 for non-interacting superparamagnetic system. Moreover, for a system having interparticle interaction or spin glass like nature, k value lies within the order of $10^{-2}$ to $10^{-3}$. Herein, k yields 0.029 for CE and 0.032 for HCE. The observed value of k is an order higher than that of reported value for canonical SG systems indicating cluster-SG type behaviour in both CE and HCE systems \cite{1, 52}.

\begin{figure}[hbt!]     
\centering           
\includegraphics[width=8.5 cm,height=12 cm]{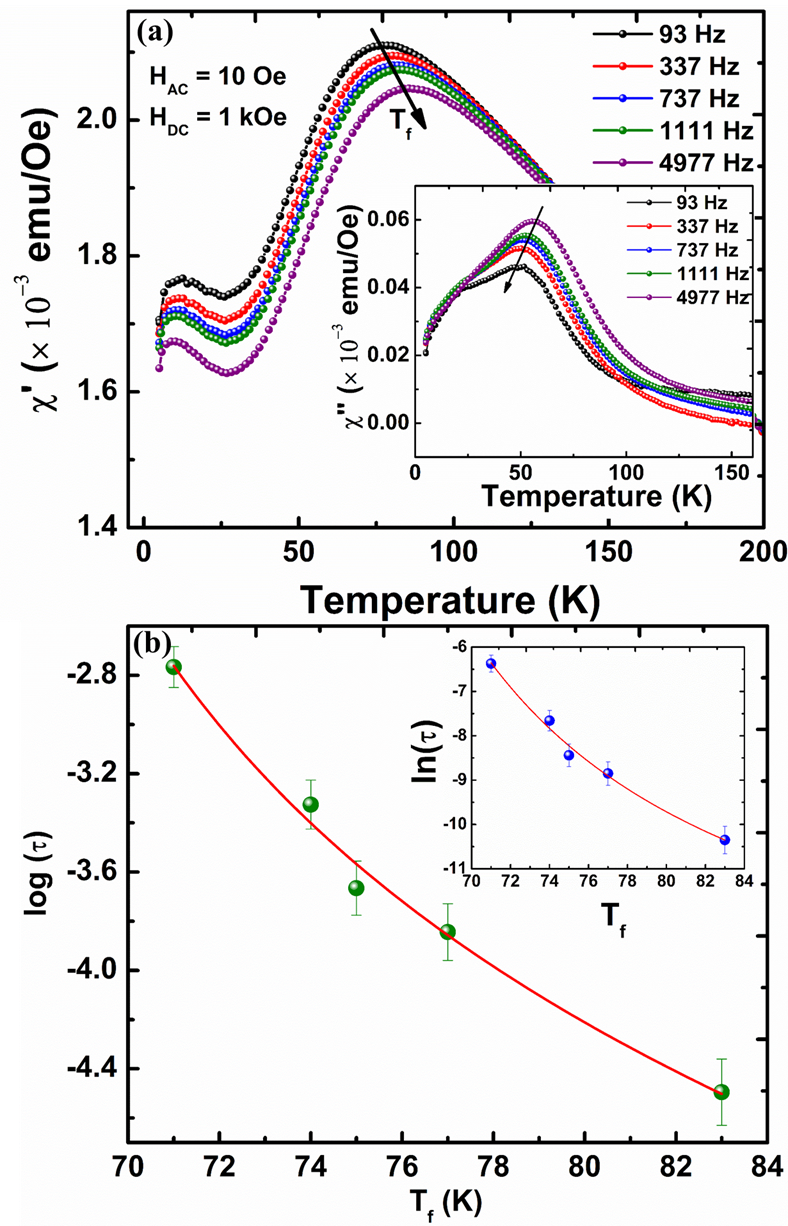}
\caption{\label{Fig:wide} (a)In-phase ac susceptibility at applied field $H_DC$ = 1 kOe with $H_AC$ = 10 Oe, out-of-phase ac susceptibility in inset; (b)log-log plot of power law, VF-law fitting in inset for system HCE}
\end{figure}

For further evaluation, Arrhenius model for non-interacting or weakly interacting particles is employed to fit frequency dependent $T_{f}$ values, which can be expressed as \cite{52},
\begin{equation}
\tau = \tau_0\;{exp\left[\frac{E_{a}}{k_BT_f}\right]},
\end{equation}

where $\tau_0$ is the relaxation time between two consecutive attempts, $E_a/k_B$ is the activation energy which measures the energy barrier required for the separation of the metastable states. The linear fitting of $\ln$ ($\tau$) versus $T_f$ curve yields an unphysical values of $\tau_0$ and $E_a/k_B$ for both the systems [$\tau_0$ = $1.74\times10^{-34}$s  and $E_a/k_B$ = 11393  K for CE; $\tau_0$ = $6.9\times10^{-32}$ s  and $E_a/k_B$ = 7627 K]. This unphysical results ensure that the spin dynamics is not only due to the flipping of the single spin, but also due to the collaborative nature of inter-particle interactions. Therefore, a dynamical scaling law is taken into consideration known as Vogel-Fulcher (VF) model by considering the contribution of interaction among the spins. VF-model can be depicted as [39],
\begin{equation}
\tau = \tau_0\;{exp\left[\frac{E_{a}}{k_B(T_f - T_0)}\right]},
\end{equation}
where, $T_0$ interprets the characteristics temperature to measure qualitatively the interparticle or intercluster interaction energy. To fit the frequency dependent $T_f$ values, the equation 9 can be rewritten as,

\begin{equation}
\ln (\tau) = \ln(\tau_0) + \frac{E_a}{k_B(T_f - T_0)},
\end{equation}

Figure shows the fitting of $\ln$ ($\tau$) versus $T_f$,  which is showing best linear fit with $\tau_0$ = $8.85\times10^{-9}$ s, $T_0$ = 142.58 K and $E_a/k_B$ = 216 K for the system CE ; $\tau_0$ = $3.88\times10^{-10}$ s, $T_0$ = 97 K and $E_a/k_B$ = 309.16 K for the system HCE. The obtained non-zero value of $T_0$ with other significant values of the parameters confirm the agreement of VF-law for both the systems and ensures the contribution of finite interaction among the spins to the spin dynamics. The observed $\tau_0$ values for both CE and HCE is differ in its magnitude by few orders from the atomic spin flipping ($\sim$ $10^{-13}$ s), but coming under the characteristic relaxation period of cluser spin glass system \cite{1, 52, 54, 55}. The system HCE shows slower relaxation of HCE than that of CE which is due to the domination of the stronger dipolar interaction among the spins. The observed activation energy for CE is $\frac{E_a}{k_B}$ $\sim$ 1.5$T_g$ and for HCE is $\frac{E_a}{k_B}$ $\sim$ 3.1$T_g$, where $T_g$ stands for the temperature for static spin glass. The system CE is showing the typical activation energy range of atomic spin glass ($\frac{E_a}{k_B}$ $<$ 2$T_g$). On the other hand, in the system HCE, slightly higher value of activation energy is observed. Such higher value in activation energy is observed in many reported cluster spin glass systems \cite{54, 55} having hollow morphology where collective freezing occurs because of superspin moments or spin clusters.
\begin{figure*}[hbt!]     
\centering           
\includegraphics[width=16.5 cm,height=10 cm]{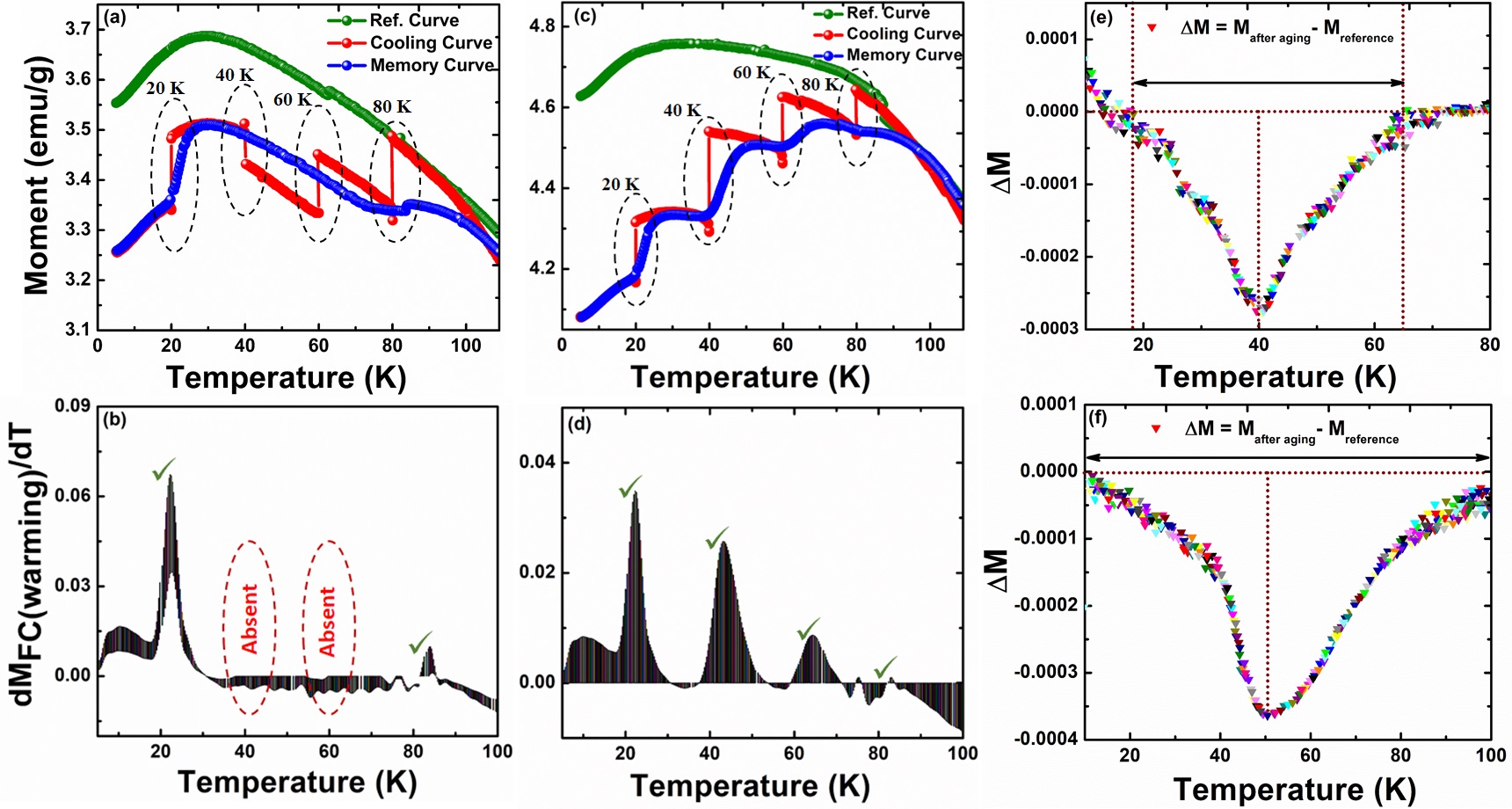}
\caption{\label{Fig:wide} FC memory effect for the system (a) CE, (c) HCE; temperature derivative of FC warming curve (b) CE, (d) HCE; ZFC memory effect (e) CE, (f) HCE}
\end{figure*}

To execute the dynamic critical slowing of spins,the frequency dependent $T_f$ values at $H_{dc}$ = 0 are fitted by power law provided by dynamic scaling theory,
\begin{equation}
\tau = \tau^*\left[\frac{T_f - T_g}{T_g}\right]^{-zv'},
\end{equation}
where, $\tau^*$ is the relaxation period for single spin flip, $T_g$ is glass transition temperature, z stands for dynamic critical exponent and $v'$ gives the critical exponent for correlation length. The correlation length can be expressed as, $\zeta$ = $(T_f/T_g - 1)^{-v'}$ and the spin relaxation time is connected to $\zeta$ as $\tau$ $\propto$ $\zeta^z$. To fit the ac susceptibility data, power law [Eq. 11] can be expressed simply as,

\begin{equation}
\log(\tau) = \log(\tau^*) - zv' \log(\left[\frac{T_f - T_g}{T_g}\right]),
\end{equation}   

The linear fitting of the $\log(\tau)$ versus $T_f$ is shown in figure 7(a) for system CE and figure 7(b) for HCE. The curves show best fitting at $\tau^*$ = $7.7\times10^{-11}$ s , $zv'= 6.2 \pm 1.1$ with glass transition temperature $T_g$ = 150.4 K for CE , and $\tau^*$ = $1.3\times10^{-10}$ s, $zv'= 6.3 \pm 0.5$ with $T_g$ = 108.9 K for HCE. The observed parameters provide some physical value to shed light on the spin dynamics. In case of a conventional spin glass system, the value of $zv'$ and $\tau^*$ typically varies from 4 to 12 and $10^{-10}$ to $10^{-13}$ s \cite{52, 37, 56}. For canonical SG system $\tau^*$ lies between the order of $10^{-12}$ to $10^{-13}$ s and for cluster SG $\tau^*$ should lie between $10^{-7}$ to $10^{-11}$ s \cite{1, 52, 39, 56}. It is observed that the system fitted data of CE and HCE fall in the range of cluster SG system. The study is performed with an applied DC field $H_{DC}$ = 1 kOe for system HCE along with the AC field. The incorporation of DC field leads to shifting of $T_f$ towards lower temperature region. A change in the fitted values of the parameters obtained from the VF-law [Eq. 10] and power law [Eq. 12]. VF-law yields $\tau_0$, $T_0$ and $E_a/k_B$ as $1.6\times10^{-7}$, 55 K and 145.5 K; power law yields $\tau^*$ = $2.8\times10^{-7}$ and $zv'=3.7$. The observed activation energy at $H_{dc}$ = 1 kOe is 2.6$T_g$ which is less than that of the activation energy calculated at $H_{dc}$ = 0. It is because, the less anisotropic cluster regions start to align leading to spin freezing at lower thermal energy due to the high DC field. Though there is a change in $\tau^*$ and $zv'$, the values are still in the range of cluster-SG phase.

It is observed that the trend of relaxation time $\tau$ in both the systems can be understood properly by validating both the dynamic scalling laws, such as Vogel-Fulcher law and power law. Although a slight difference is observed in $\tau^*$ value analyzed from power law than that of $\tau_0$ calculated from VF law, such kind of behavior is evident by many reported cluster SG-systems \cite{1, 52, 55, 38, 39, 40}. Further, it is observed that the value of $T_g$ is higher than $T_0$ by a very less fractional value following the trend of cluster-SG systems \cite{52}. However, for further investigation, the Tholence criterion $\delta$T$_{Th}$ = $\frac{T_f - T_0}{T_f}$ is employed \cite {57}. The values of $\delta$T$_{Th}$ is obtained as $\sim$ 0.1 [taking $T_f$ $\simeq$ 160.8 K and $T_0$ $\simeq$ 142.58 K] for CE and $\sim$ 0.2 [taking $T_f$ $\simeq$ 117 K and $T_0$ $\simeq$ 97 K] for HCE at zero dc field are validating the reported value for cluster-SG systems \cite{52}.

\subsection{\textbf{Nonequilibrium dynamics}}
\subsubsection{\textbf{Magnetic memory effect}}
To ensure the existence of nonergodicity and spin dynamics below blocking temperature, magnetic memory effect study is performed using both Field-Cooling (FC) and Zero Field Cooling (ZFC) conditions \cite{18, 19, 26}. The FC memory effect results are shown in Fig. 9 (a), 9 (b), 9(c) and 9 (d), and ZFC memory effect results are shown in figure 9(e) and 9(f) respectively. During FC condition, initially temperature dependent magnetization protocol is followed to obtain the reference curve. Later on, the systems are cooled upto 5 K with a cooling rate of 1 K/min in a magnetic field of 100 Oe. Few intermittance stops are employed at 80 K, 60 K, 40 K and 20 K for waiting duration 1 hour each. At each stopping temperature, when the probe field is turned off during waiting time, the magnetic moments get relaxed. Because of moment relaxation, the magnetization decreases after each stoppage with respect to the reference curve. A prominent step-like behaviour is noticed in such procedure at each stop. After each stopping duration, the probe field is turned on and FC process is recommenced. The recorded cooling curve with the intermittance stoppage is marked as cooling curve. Once the system reaches at temperature 5K, the magnetization is measured again following the warming process up to room temperature with the applied field without any intermittance interruption. The recorded warming curve is marked as memory curve. If the systems will able to memorise the spin identification left throughout the cooling process, the magnetic memory is said to exist in the systems. The system CE is able to memorize two pronounced steps at 20 K and at 80 K. In contrast, the system HCE is manifesting distinctly four steps of memory. Differentiation of the memory curves with respect to the temperature also provides the imprint of memory step at each halted temperature for firm confirmation. The observed steps ensure that system retrieve energy configuration which is marked by energy barrier redistribution through cooling process. CE having lower interparticle interaction could recover only lower energy magnetic arrangement.

A spin glass system exhibits non-equilibrium nature as very large period is required to attain equilibrium magnetization below spin glass critical temperature \cite{21, 22, 27}. The ZFC memory effect is examined at 40 K at 50 Oe applied field. Initially normal ZFC protocol for magnetization vs. temperature is performed at 50 Oe for the reference curve. After that, both the systems are cooled down to a certain low temperature, 40 K in absence of any magnetic field and aged for $10^4$ seconds. After the aging, the system is further cooled to 5 K. At this stage, warming process is continued with a field of 50 Oe and the moment is measured up-to room temperature. The obtained warming curve after aging is marked as $M_{memory,ZFC}$.A comparison between memory curve and ZFC reference curve shows a prominent dip at the interrupted region in both the systems as shown in Fig. 9(e, f). It provides evidence of spin glassy state in systems as moment dynamics has slowed down below a certain temperature. The observed non-zero moment $\Delta M$ in between temperature range $\sim$ 10 K to 50 K for CE and $\sim$ 10 K to 100 K for HCE ensure that systems get relaxed towards steady dynamics during foisted waiting time, as explained in both spin glass models, hierarchical energy model \cite{58} and spin droplet model \cite{59}. In case of the droplet model, the excitation of spin glass configure compact domains and non-equilibrium behaviour of spin dynamics increases the volume of droplet with time. During aging, as temperature becomes constant, growth of droplet and frozen energy barrier associated with it occurs simultaneously due to the absence of perturbation. It can be recovered once warming starts. At interrupted temperature, adequately low energy barriers result in flipping of thermally energetic cluster upon warming and provides low magnetization moments during memory path M$_{mem,ZFC}$ over reference curve M$_{ref,ZFC}$.

\begin{figure*}[hbt!]     
\centering           
\includegraphics[width=13.5 cm,height=10 cm]{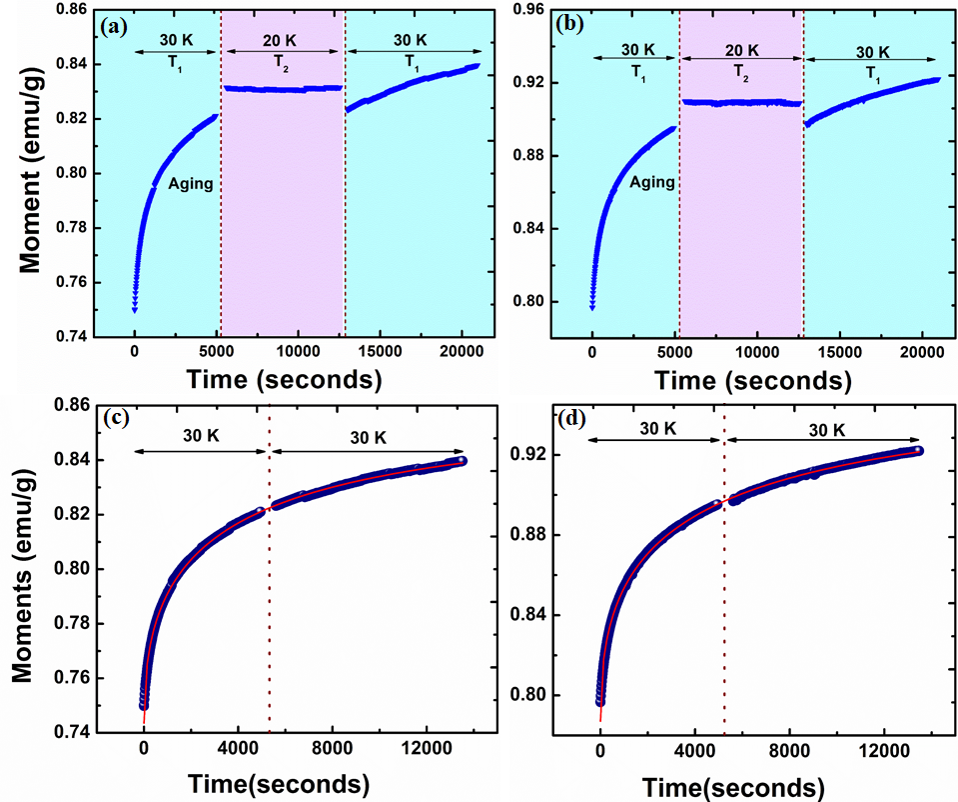}
\caption{\label{Fig:wide} Magnetic relaxation measurements using ZFC protocol for (a)CE and (b) HCE, Continuation of relaxation trend for (c) CE and (d) HCE}
\end{figure*}

We further study magnetic relaxation using ZFC protocol as shown in Fig. 10 \cite{17}. The systems are cooled to temperature 30 K (T$_1$) and aged for a period of 5000 seconds (t$_1$) at applied field 50 Oe. Aging can be observed as a significance of jagged nature of disordered landscape of spins. This conforms to the crossing of thermally activated energy barriers leading to slow relaxation of the spins towards minimum energy levels. As the moments unable to attain its equilibrium state, they start to relax very slowly towards the direction of the applied field and follows a logarithmic trend as shown in Fig. 10 (a, b) marked as aging. Further, the systems are temporary cooled down to 20 K (T$_2$) and the moment are measured for the period of 7000 seconds (t$_2$). At this step the spin dynamics is not following the trend of T$_1$, but the moments are arrested and become constant during this entire period. The frozen spins start to adjust at T$_2$ and refuse to slow down with free energy barriers. When temperature comes to T$_1$, moments recommence its ascending relaxation trend from preceding value. The systems are able to memorize strongly its age at T$_1$. The continuity of moment relaxation is observed in joined curves (shown in Fig. 10(c, d)) signifying the memory effect in cluster spin-glass systems and complete freezing dynamics between 20 K and 30 K in both the systems. The non-compact spin clusters are large enough and it cannot be frozen at high temperature but left the surrounding spins comparatively free. As temperature starts to increase, small-small cluster of spins begin to freeze resulting an aging signal, whereas large spin cluster begin to be blocked completely resulting memory effect \cite{25,26}.

The continuity curves of both the systems presented in Fig. 10 (c) and 10 (d) are fitted with stretched exponential function,

\begin{equation}
M(t) = M_0 - M_g exp\left[-\left(\frac{t}{\tau}\right)^{\beta}\right],
\end{equation}

where $M_O$ and $M_g$ stand for intrinsic magnetization and magnetization for glassy component respectively. $\tau$ represents characteristic relaxation time constant and $\beta$ is the function of temperature known as stretching exponents which value lies between 0 and 1. $\beta$ signifies the spin dynamics and the energy barrier distribution involed in spin relaxation. If the fitting of equation 13 yields $\beta$ = 0, it means there is no relaxation of spins. If $\beta$ gives a value of unity, it indicates that the relaxation of system occurs with only single time constant.  Moreover, systems having uniform energy barrier show $\beta$ = 1. Although, 0$<$ $\beta$ $<$ 1 signifies the presence of distribution of energy barriers in the system. The obtained fitted value of $\beta$ for the system CE and HCE are 0.47 and 0.45 respectively and these values are lying under the reported values of various glassy systems \cite{37, 56, 17}. Moreover, as the obtained values of $\beta$ are less than 1, it manifests that the systems undergo multiple intermediate metastable states \cite{52}. Therefore, the activation energy involeved in such systems has to overcome multiple anisotropic energy barriers.

In order to explain the contribution of individual factor affecting the magnetic properties of the two systems, the enenrgy expression \cite{60} can be introduced in a generalized way introducing the surface driven contributions,

\begin{equation}
\begin{split}
E = - {\mu_B M \sum S} - \frac{1}{2} [V_{net} \{ \sum S_i S_{i+1} cos \theta_i\\ + \sum {S_j} S_ {j+1} cos \theta_j\}] - K_{ani} ^ {outer} {\sum S_i sin^2 (\theta_i - \varphi _i)} \\ - K_{ani} ^ {inner} {\sum S_j sin^2 (\theta_j - \varphi _j)},
\end{split}
\end{equation}

Here the first term provides the Zeeman energy contribution to the energy density when a magnetic field is applied. As the pinned surface spins are dominated by the dipolar interaction, the second term is introduced for the dipolar interaction contribution among the disordered spins of the surface with dipolar coupling constant $V_{net}$. A new term is introduced for the presence of additional surface which enhances the dipolar contribution. As the demagnetizing field among the spins is dominating the interaction among the spins, the crystalline field developed a strong uniaxial anisotropy leading to a anistropy term in the energy expression. For such an anisotropic dipolar system, the anisotropy contribution is represented in the third and fourth term of the equation. If we consider the energy expression for the system CE, the later part of the dipolar contribution term and the fourth term of the equation 14 will not be considered. Moreover, for the system HCE, all the contributions will be present for the energy expression.

The interparticle interaction plays a dominant role to produce a complex free-energy landscape with higher degree of freedom. Thus, presence of memory features is a consequence of frustration arising due to the competition between disordered spins and crystallographic anisotropy which depends not only on the shape, size, compactness and interparticle interaction of the nanoparticles, but also the relative geometry organization of the magnetic ensembles. The obtained outcomes can be understood by considering an approach, which is based on the degree of disordered surface spins with varied strength of dipolar interaction along with varied spatial arrangement of the easy axes leading to modulation in collective magnetic behaviour. The anisotropic nanoparticles of system CE bearing uniaxial anisotropy are arranged with an optimum interparticle space resulting less dipolar strength. The presence of space among each particles allow rotation of individual spin resulting less complex energy landscape. On the contrary, despite of having higher demagnetization effect, the system HCE shows higher coercivity, reduced blocking temperature and more pronounced magnetic memory effect over CE which is not supporting both the DBF model and MT model. In addition to higher interaction, presence of hollow interior in HCE provides enhanced surface effects like large surface to volume ratio with lowest energy domains configuration along with an additional surface from the interior side having higher degree of pinned spins. The higher degree of frustrated surface spins with complex anisotropic barriers and closer spatial arrangement of anisotropic nanoparticles with a hollow geometry provide wider $\Delta$M range in ZFC memory effect and more prominent FC memory effect.

\section{\textbf{SUMMARY}}
In summary, we develop two differently organized identical systems CE and HCE by tuning their geometry and investigate systematically their dynamic magnetic nature. The presence of pure phase of ZnFe$_2$O$_4$ is evident from powder XRD and the organization pattern of anisotropic nanoparticles in the ensemble is ensured by SAXS, SANS and HRTEM. The domination of demagnetization interaction with primary evidence of SG transition at low temperature is confirmed in both the systems from dc magnetization study. ac susceptibility analysis is performed for further confirmation of SG transition. The fitted parameters obtained from dynamic scaling laws, Mydosh parameter and Tholence criterion value are consistent with the cluster-SG systems for both the systems. An increment in activation energy is observed at HCE, $E_a/k_B$ $\simeq$ 309.1 K than that of CE, $E_a/k_B$ $\simeq$ 216 K. Both FC/ZFC memory effects and ZFC spin relaxation process in negative temperature cycle infer presence of cluster-SG state with establishment of spin frozen state. Moreover, HCE having higher demagnetization strength is exhibiting all prominent FC memory and wider non-zero range of $\Delta$M in ZFC memory effect than that of CE. It is due to its highly competing and frustrated surface spins in addition to its hollow core which enhance the surface driven effects. However, the observed unusual enhancement in coercivity and reduced blocking temperature with enhanced demagnetizing interaction is a consequence of complex anisotropy energy barrier due to interacting energy landscape. This study demonstrates the slower spin relaxation dynamics and the enhancement in magnetic memory effect by simply tuning the demagnetization interaction and respective geometry of the ensembles.

\subsection{\textbf{ACKNOWLEDGEMENT}}
The authors acknowledge the UGC-DAE-CSR, Mumbai Centre, India, for providing financial support vide grant no. UDCSR/MUM/AO/CSR-M-249/2017. The authors acknowledge Innovation and research grant 2019, Tezpur University, India having vide no. DoRD/RIG/10-73/1362-A Dated 19/02/2019. KK would like to acknowledge Dr. Koushik Saikia for fruitful discussion during manuscript development.

\nocite{*}\
\bibliography{main}
\end{document}